\documentclass[fleqn,usenatbib]{mnras}

\usepackage{graphicx}
\usepackage{subcaption}
\usepackage{booktabs}
\usepackage{amsmath}
\usepackage{lipsum}
\usepackage{comment}
\usepackage{adjustbox}
\usepackage{multirow}
\usepackage{txfonts}
\usepackage{float}
\usepackage{natbib}
\usepackage{hyperref}
\usepackage{lscape}
\usepackage{xspace}
\usepackage{xargs}
\usepackage{enumitem}
\newcommand{\ha}{\textup{H$\alpha$}\xspace}
\newcommand{\hb}{\textup{H$\beta$}\xspace}
\newcommandx{\hei}[1][1=]{\textup{He\,\textsc{i}}{#1}\xspace}
\newcommand{\pab}{\textup{Pa$\beta$}\xspace}
\newcommand{\pag}{\textup{Pa$\gamma$}\xspace}
\newcommand{\oiii}{\textup{[O\,\textsc{iii}]}\xspace}
\newcommand{\nii}{\textup{[N\,\textsc{ii}]}\xspace}
\newcommand{\kms}{\ensuremath{\mathrm{km\,s^{-1}}}\xspace}
\defcitealias{Juodzbalis2024a}{J24}
\defcitealias{rusakov2025}{R25}

\usepackage[T1]{fontenc}
\DeclareRobustCommand{\VAN}[3]{#2}
\let\VANthebibliography\thebibliography
\def\thebibliography{\DeclareRobustCommand{\VAN}[3]{##3}\VANthebibliography}

\title[Ruling out electron scattering in Rosetta Stone]{Ruling out  dominant electron scattering in Little Red Dots' \emph{Rosetta Stone} using multiple hydrogen lines}
   
\author[Brazzini, M. et al.]{\parbox{\textwidth}{
Matilde Brazzini$^{1,2,3,4}$ \thanks{E-mail: matilde.brazzini@inaf.it},
Francesco D'Eugenio$^{1,2}$ \thanks{E-mail: fd391@cam.ac.uk},
Roberto Maiolino$^{1,2,5}$,
Ignas Juod{\v z}balis$^{1,2}$,
Xihan Ji$^{1,2}$,
and Jan Scholtz$^{1,2}$
}\vspace{0.4cm}
\\
\parbox{\textwidth}{
$^{1}$Kavli Institute for Cosmology, University of Cambridge, Madingley Road, Cambridge, CB3 0HA, United Kingdom\\
$^{2}$Cavendish Laboratory - Department of Physics, University of Cambridge, 19 JJ Thomson Avenue, Cambridge, CB3 0HE, United Kingdom \\
$^{3}$Department of Physics, Astronomy Section, University of Trieste, Via G.B. Tiepolo, 11, I-34143 Trieste, Italy\\ 
$^{4}$INAF - Osservatorio Astronomico di Trieste, Via G. B. Tiepolo 11, I-34143 Trieste, Italy\\ 
$^{5}$Department of Physics and Astronomy, University College London, Gower Street, London WC1E 6BT, United Kingdom
}}

\date{Accepted XXX. Received YYY; in original form ZZZ}
\pubyear{\the\year{}}

\begin{document}
\label{firstpage}
\pagerange{\pageref{firstpage}--\pageref{lastpage}}
\maketitle

\begin{abstract}
 The majority of Little Red Dots (LRDs) hosting Active Galactic Nuclei (AGN) exhibits broad H$\alpha$ emission, which recent studies propose originates from scattering off free electrons within an ionized and dense medium embedding the Broad Line Region (BLR), rather than directly from the BLR itself. 
 This model suggests that the observed broad lines may be intrinsically narrower than observed, which  would lead to black hole masses that are up to two orders of magnitude smaller than what inferred
 when assuming that the whole broad line comes from the BLR.
 To test this model, we present a joint analysis of multiple hydrogen recombination lines in the ``Rosetta Stone'' AGN, the brightest known LRD at $z$=2.26.
 We show that \ha, \hb and \pab have different spectral profiles, which is inconsistent with the predictions of the simple electron scattering scenario. 
 Additionally,
 we test a variety of exponential models and show that none of them can simultaneously reproduce all three line profiles with physically plausible parameters. 
 The inadequacy of these models for the Rosetta Stone implies that the scenario of electron scattering by an ionized medium surrounding the BLR is not universally applicable to LRDs and AGN, and therefore provides a counterexample to the claim of a universal and systematic overestimation of black hole masses.

\end{abstract}

%
\section{Introduction}
James Webb Space Telescope (\textit{JWST}) has revolutionised our understanding of supermassive black hole (SMBH) formation and evolution in the early Universe, revealing a new population of low-mass, low-luminosity active galactic nuclei \citep[AGN;][]{harikane+2023, Kocevski+2023, Maiolino+2024,  Matthee+2024, juodzbalis2025}. 
These sources are usually compact, and weak in both X-rays and radio emission \citep{Mazzolari+2024a, Mazzolari+2024b, Maiolino2025}, leading to initial scepticism regarding their AGN nature.  
However, their classification is now firmly established through their time variability \citep{ji2025, furtak+2025}, and the detection of broad \ha emission (full-width at half maximum, FWHM~$\gtrsim1000~\kms$) with no broad counterpart in \oiii$\lambda\lambda$4959,5007, ruling out outflows and confirming that broad \ha originates from the broad line region (BLR). 
A fraction of these AGN (ranging from 10-30\%, \citealp{Hainline+2025}, to $\sim 44\%$, \citealp{zhang+2025}) further exhibits unusual properties compared to standard Type I, including a characteristic `v'-shaped UV-to-optical spectral energy distribution (SED), and a prominent spectral break at rest-frame wavelength $\sim 3700$\AA\ (i.e., the Balmer break). 
These sources have been dubbed little red dots (LRDs). 

The discovery of LRDs has raised more questions than current observations can answer.
In particular, an increasing number of high-redshift LRDs exhibits deep Balmer and \hei$\lambda 10829$ absorption features \citep{Matthee+2024, Juodzbalis2024a, Wang+2025, ji2025, loiacono2025, deugenio+2025_lrd_2, deugenio+2025_lrd}, whose physical origin remains uncertain. 
One plausible explanation invokes absorption by extremely dense gas embedding the BLR ($n_\text{H} \gtrsim 10^9$ cm$^{-3}$, \citealp{inayoshi+2025}). 
If this interpretation is correct, the presence of such high-density gas could also account for the X-ray weakness of these sources --assuming they are Compton-thick-- and also for their prominent Balmer break, ruling out its stellar origin \citep{Labbe+2024}. 
Moreover, if the medium is ionised, it is possible that the observed broad emission line widths may arise from electron scattering within this gaseous envelope, rather than from bulk gas motion in the BLR itself. 
This hypothesis, first tested by \citet[][hereafter: \citetalias{rusakov2025}]{rusakov2025}, suggests that the intrinsic width of the broad lines could be 5--10 times narrower than currently measured, potentially leading to systematic overestimates of black hole masses by up to two orders of magnitude. 
In particular, \citetalias{rusakov2025} studied the \ha line in a sample of 13 JWST-discovered AGN at $z\sim 3\text{--}7$ modelling the broad wings with an exponential profile, and compared it with a single-Gaussian broad-line model. They found that the former is statistically preferred for all but one of the sources in their sample. 
In this work, we aim to further explore the electron scattering scenario by studying the line profiles of multiple hydrogen recombination lines in a luminous LRD at cosmic noon, GN-28074, that, given its brightness, and prominent emission and absorption features, has been considered the ``Rosetta Stone'' of Little Red Dots \citep{Juodzbalis2024a}.
Our specific target is presented in Section \ref{section-data}, and its peculiar line emission is modelled in Section \ref{section-emission-line-modelling}. 
We discuss our results in Section \ref{section-discussion}, and draw our conclusions in Section \ref{section-conclusions}.
Throughout this work, we assume a flat $\Lambda$CDM cosmology from \cite{plank}. 

\section{Target and data}
\label{section-data}

GN-28074 is a broad-line AGN at $z=2.26$ in the GOODS-N field (ra=189.06458 deg, dec=62.2738 deg), observed with JWST as part of the JWST Advanced Deep Extragalactic Survey \citep[JADES;][Program ID 1181]{Eisenstein2023}.
The AGN has bolometric luminosity of $\log (L_{\text{bol}})=45.76$ \citep{juodzbalis2025}. Undetected in X-rays, it is has currently the lowest X-ray to bolometric emission ratio among high-z AGN ($L_X/L_{\text{bol}}< 10^{-4}$), three orders of magnitude lower than what is observed in local AGN of similar luminosity \citep{Maiolino2025}.
It displays the characteristic features of LRDs listed above: a `v'-shaped SED, broad lines, a Balmer break, and is spatially unresolved at red wavelengths.
It also shows deep blueshifted absorption in \ha, \hb\ and \hei \citep{Juodzbalis2024a}. 
All these features make GN-28074 the ideal object to test the electron scattering scenario, especially given its brightness and redshift, which allow \ha, \hb and \pab lines to be simultaneously observed by \textit{JWST} at high signal-to-noise ratio (SNR). 
We note that \pag is also within the observed spectral range, but we neglect it in the following analysis due to strong blending with the meta-stable \hei line.
In this work, we focus on the medium-resolution spectroscopy ($R=700\text{--}1500$), obtained with 1.7 hours of exposure using the NIRSpec \citep{jakobsen2022} micro-shutter assembly 
\citep{ferruit+2022}.
These data have been processed following the JADES data reduction pipeline \citep{Bunker2024, deugenio2025_jadesdr3}, with additional updates described in \cite{Juodzbalis2024a}. 

\begin{figure}
    \centering
    \includegraphics[width=0.9\linewidth]{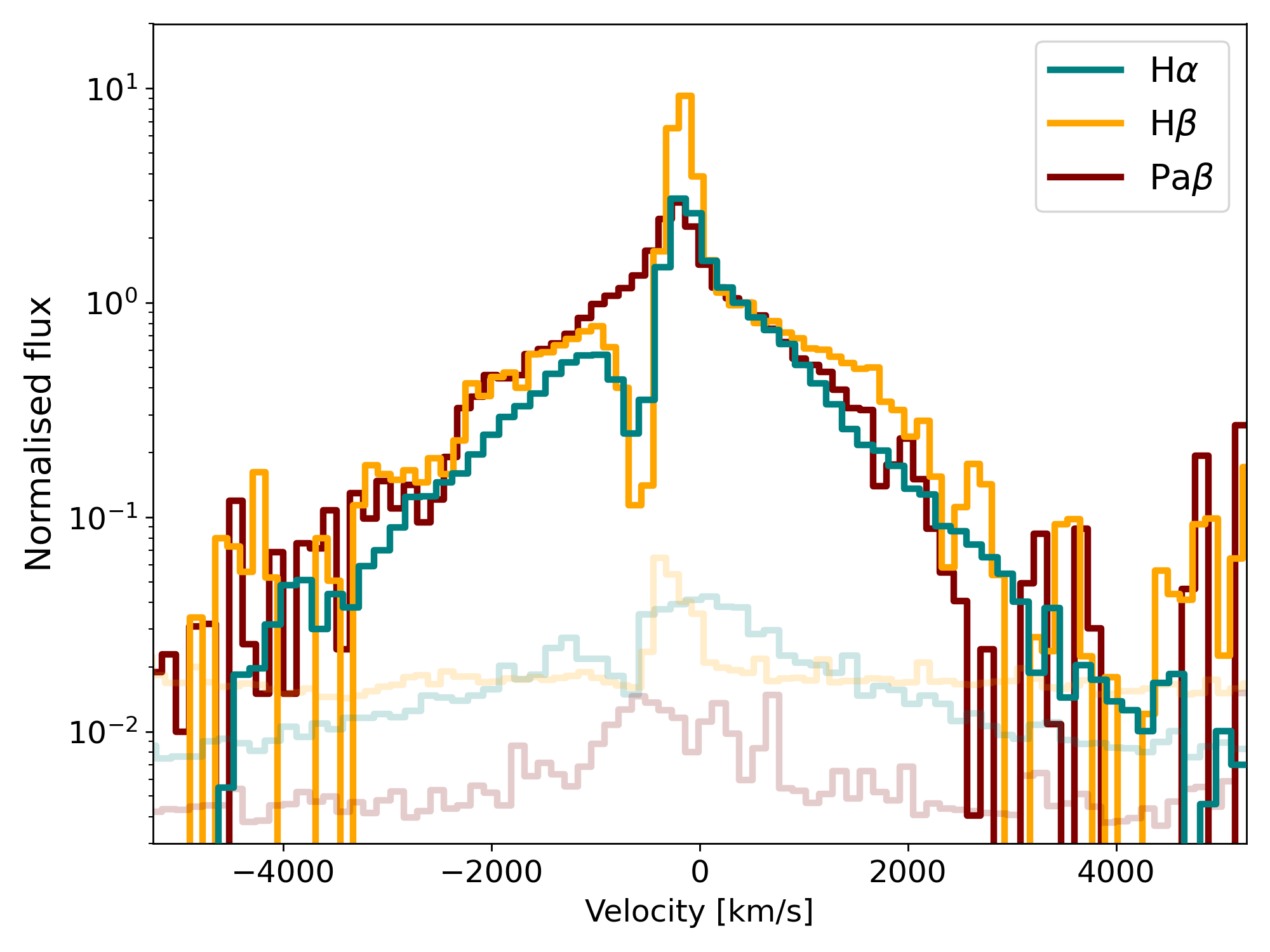}
    \caption{Line profile comparison in velocity space and on logarithmic scale for \ha, \hb\ and \pab. Their fluxes are normalised to the red 500 \kms. Spectral errors are reported in lighter colours. }
    \label{fig:line-profiles-vel-space}
\end{figure}

\section{Emission line profiles}
\label{section-emission-line-modelling}


The electron scattering scenario assumes that the BLR is embedded in a high column density medium of ionized gas, whose scattering produces exponential wings in the emission lines. 
Given that electron scattering is wavelength independent, at least over the spectral range considered in this work, all lines from the BLR should be scattered in the same profile. 
In Fig.\ref{fig:line-profiles-vel-space} we compare (in logarithmic plot) \ha, \hb and \pab profiles, normalized to the red wing of the line at 500~\kms. 
Clearly, regardless of the blueshifted absorption by neutral gas, the wings of the three lines have different shapes, either on the blue or on the red side, challenging the idea of electron scattering as the dominant mechanism for line broadening. 
To investigate this further, we fit emission line profiles using a variety of models, as detailed below.
Line fitting is performed within a Bayesian framework, adopting flat, non-informative priors unless otherwise specified. 
Posterior probability distributions of all parameters are estimated using the Markov-Chain Monte-Carlo integrator \textsc{emcee} \citep{emcee_ref}, with the chains initialised near the least-squares solution from the \textsc{scipy}'s \textsc{curve\_fit} function. 

We model the continuum around emission lines with a linear function. Each hydrogen line is fit with a combination of three Gaussians plus an exponential, representing, respectively, the narrow component, an outflow (seen in \oiii), the broad unscattered and the broad scattered emission from the BLR.
We allow a sub-pixel velocity offset of $ \pm 100$ \kms between the three different hydrogen lines, to accommodate possible wavelength calibration issues between the different NIRSpec gratings \citep{deugenio2025_jadesdr3}.  
The narrow Gaussian components are forced to have the same intrinsic width, while we put no constraints between the broad line widths. 
We allow for a velocity shift between the narrow and broad components in the range $\pm 500$~\kms. 
The velocity and velocity dispersion of the outflow Gaussian components are fixed to those of \oiii, which is fitted separately (Appendix \ref{appendix-O3-fit}). 
We impose the line widths to vary, respectively, between 0 and 700 \kms for the narrow, and 1000 and 10,000 \kms for the broad components. 
For the outflow fitting, we let the line width vary in the range 100 -- 10,000 \kms. 
Lastly, the exponential term representing the scattered light from the BLR is modelled by convolving the BLR Gaussian profile, $G_\text{BLR}(\lambda)$, with a symmetric exponential \citep{laor2006} in the form: 
\begin{equation}
    E(\lambda_0, W; \lambda) \propto e^{- \frac{|\lambda - \lambda_0 |}{W}},
\end{equation}
where $\lambda_0$ is the central wavelength  (assumed to be the same of the Gaussian BLR), and $W$ is the exponential width, which is allowed to vary in the range 500-10,000 \kms. 
The full BLR emission for each line is therefore given by: 
\begin{equation}
    f_\text{scatt} E(\lambda)*G_\text{BLR}(\lambda) + (1 - f_\text{scatt}) G_\text{BLR}(\lambda),
\end{equation}
where $f_\text{scatt}$ is the fraction of scattered light.
We allow $W$ and $f_\text{scatt}$ to vary independently between the three hydrogen lines.

Balmer absorption is modelled using a standard attenuation model \citep{Juodzbalis2024a, deugenio2025_jadesdr3}:
\begin{equation}
    I(\lambda)/I_0(\lambda) = 1 - C_f (1- e^{-\tau(\lambda)}),
\end{equation}
where $I_0(\lambda)$ is the spectral flux density before absorption, $C_f$ is the covering factor of the absorber, and $\tau(\lambda)$ is the optical depth profile, assumed to be Gaussian.
During the fit, we force the two Balmer absorbers to have the same $C_f$ and kinematics. 
$I_0 (\lambda)$ comprises the continuum and BLR emission, including both the unscattered Gaussian and the scattered exponential terms. 
Absorbing the continuum is justified by the observation of variable optical continuum in LRDs \citep{ji2025}, consistent with direct or re-processed light from the accretion disc. 
Additionally, absorbing the BLR emission is necessary to avoid unphysical negative fluxes of the absorption when considering only the continuum affected by the absorption \citep{Juodzbalis2024a}. 
This physical model implies that the absorbing medium must be located within or just outside the BLR.
All line widths are convolved with the instrument line spread function \citep[LSF;][]{jakobsen2022}, corrected by a multiplicative factor of 0.7 to take into account slit underfill \citep[e.g.,][]{greene+2024}.

\begin{figure*}
    \centering
    \includegraphics[width=0.9\linewidth]{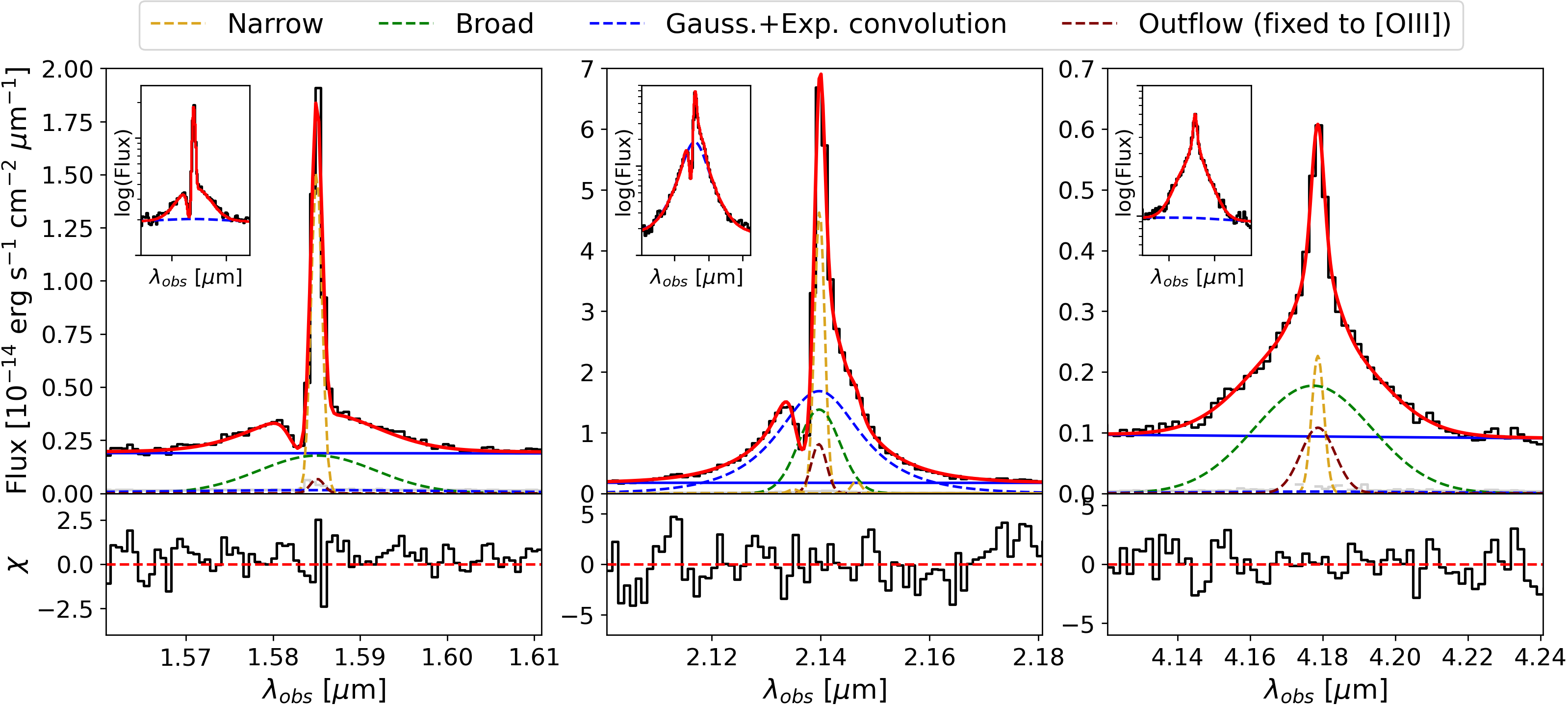}
    \caption{Best-fit models for \hb\ (left), \ha (centre) and \pab\ (right). Individual model components are reported in different colours; each line is decomposed in three Gaussians (narrow, outflow and broad unscattered components) and the exponential convolution representing the broad scattered component. For the \hb and \ha, the absorber is also modelled. The three inset panels report the line profiles in logarithmic scale, with the best-fit model in red and the exponential component in blue. 
    }
    \label{fig:line-fitting-results}
\end{figure*}

As a first step, we carry out the \oiii\ line fit to characterise the outflow component. 
This step is crucial as the bright BLR emission prevents us from reliably decomposing the outflow properties from the Balmer lines.
In a successive step, we confirmed that including an outflow component in the hydrogen lines is necessary, as models without it are statistically disfavoured ($\Delta$BIC=25). 
We then fit the \ha line individually, and verify that the exponential model performs significantly better than the single-Gaussian model \citetext{$\Delta$BIC=2235, in agreement with \citetalias{rusakov2025}} and even the double-Gaussian model ($\Delta$BIC=72).
Finally, we carry out the simultaneous fit of \ha, \hb and \pab, assuming the baseline exponential model as presented above.
The fitting results are displayed in Figs. \ref{fig:line-fitting-results} and \ref{fig:corner-plot-exponential-fit}; the salient best-fit parameters used in the discussion are reported in Table \ref{table-baseline-model-results} and in their entirety in Appendix \ref{appendix-all-baseline-fitted-params}. 
To test for potential bias introduced by the high \ha SNR (SNR$=150$ for \ha, versus SNR$=25$ and $50$ for \hb and \pab, respectively), we separately fit the \hb line, which has the lowest SNR, and check for parameter consistency.  

\section{Discussion}
\label{section-discussion}

If the electron-scattering scenario were correct, we would expect the following properties to hold for the three hydrogen lines analysed in this work: 
\begin{enumerate}[label=\roman*.]
    \item The widths of the broad Gaussian components should be similar \citep[within 25\%;][]{GreeneHo2005}, as they represent recombination or collisionally excited emission from similar regions within the same BLR. 
    \item The exponential widths, $W$, should be the same for all lines, as electron scattering is wavelength-independent over the considered spectral range. 
    \item The observed scattered fractions $f_\text{scatt}$ should satisfy:
    \begin{equation}
    f_\text{scatt,\pab} \geq f_\text{scatt,\ha} \geq f_\text{scatt,\hb}.
    \label{eq-relation-fscatt}
\end{equation}
\end{enumerate}
The latter inequality accounts for the possible presence of dust in the scattering medium. 
If dust is present, since scattered light travels through more material than the transmitted counterpart of the same line, it is subject to higher optical depth and is therefore more likely to be absorbed. 
We thus expect that for each line, the scattered component can be more strongly attenuated than the transmitted component. This effect is stronger for bluer lines, since dust attenuation is stronger at shorter wavelengths. For this reason, dust causes the observed scattered fraction to be lower for bluer lines, even if the intrinsic scattered fraction is the same.

\begin{table}
  \caption[]{Specific line fitting results for the different analysed BLR emission models of \ha, \hb\ and \pab.  }
  \label{table-baseline-model-results}
  \centering
  \begin{adjustbox}{width=\linewidth}
  \begin{tabular}{ccccc}
    \toprule
    Model & Line & FWHM & $W$ & $f_\text{scatt}$ \\
          &      & [\kms] & [\kms] & \\
    \midrule
    \multirow{3}{*}{\parbox{3cm}{\centering Baseline\\(model 1)}} & \ha  & $1258_{-34}^{+31}$  & $1110_{-10}^{+10}$   & $0.738_{-0.15}^{+0.12}$ \\[6pt] 
                              & \hb  & $2902_{-246}^{+230}$ & $4688_{-2330}^{+2105}$ & $0.25_{-0.07}^{+0.13}$ \\[6pt]
                              & \pab & $2687_{-87}^{+72}$  & $1424_{-796}^{+2963}$ & $0.07_{-0.05}^{+0.16}$ \\
    \midrule

    \multirow{3}{*}{\parbox{3cm}{\centering $W$, $f_\text{scatt}$ free\\ FWHM$_\text{broad}$ bound\\ (model 2)}} & \ha  &   & $1152_{-13}^{+14}$   & $0.685_{-0.014}^{+0.013}$ \\[6pt] 
                              & \hb  & $1395_{-41}^{+42}$ & $1299_{-76}^{+67}$ & $>0.97$ \\[6pt]
                              & \pab &   & $811_{-23}^{+23}$ & $>0.98$ \\

    \midrule
    \multirow{3}{*}{\parbox{3cm}{\centering $W$, $f_\text{scatt}$ bound\\ FWHM$_\text{broad}$ free \\(model 3)}} & \ha  & $1217_{-75}^{+85}$  &   & \\[6pt] 
                              & \hb  & $2134_{-159}^{+167}$ & $1102_{-20}^{+20}$ & $0.74_{-0.02}^{+0.02}$  \\[6pt]
                              & \pab & $2369_{-46}^{+47}$  &  &  \\

    \bottomrule
  \end{tabular}
  \end{adjustbox}
\end{table}

\begin{figure}
    \centering
    \includegraphics[width=0.8\linewidth]{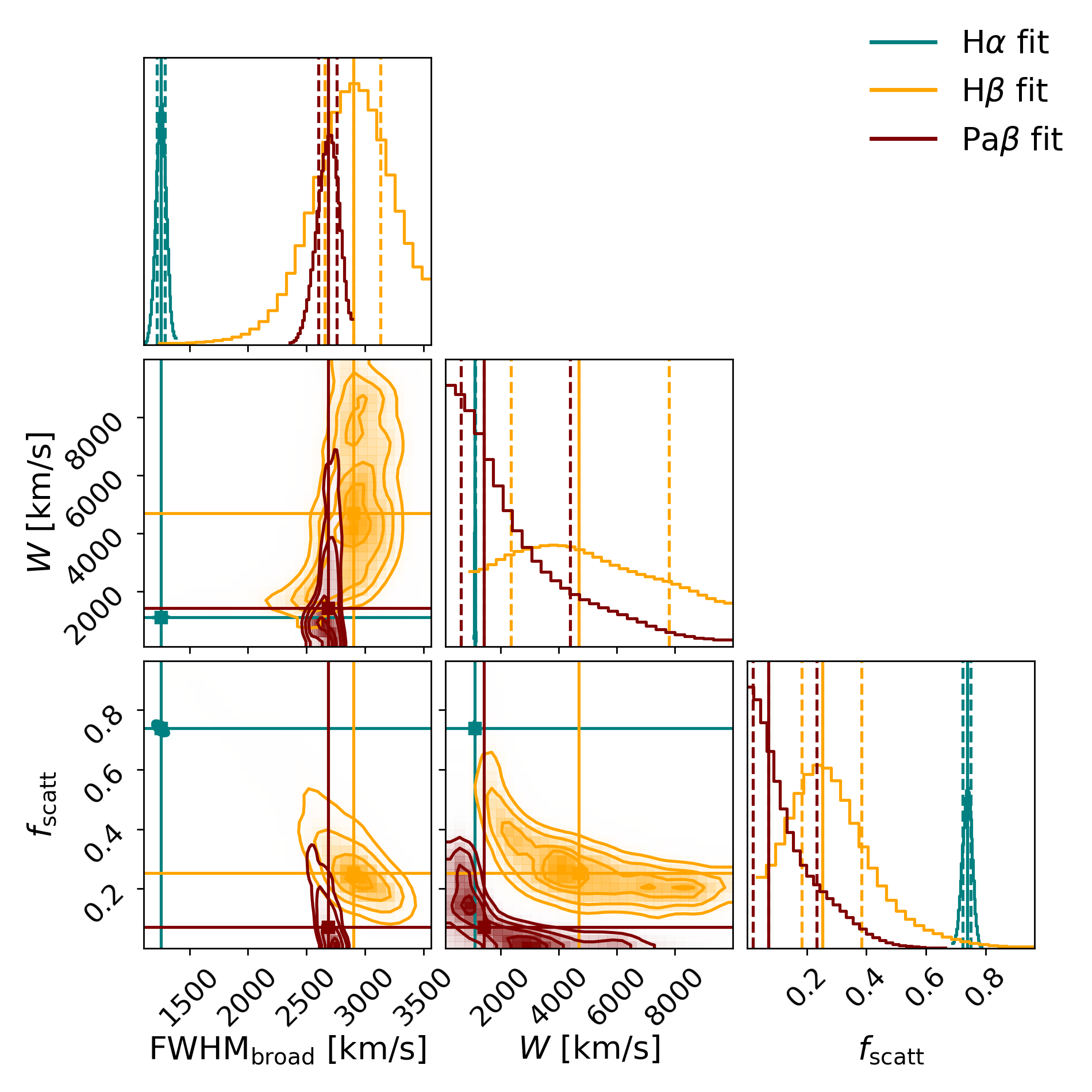}
    \caption{Posterior distributions of FWHM$_\text{broad}$, $W$ and $f_\text{scatt}$ for \ha\ (teal), \hb\ (orange), and \pab\ (maroon). 
    }
    \label{fig:corner-plot-exponential-fit}
\end{figure}


Our baseline (model 1) best-fit results for FWHM$_\text{broad}$, $W$ and $f_\text{scatt}$ are reported in Table \ref{table-baseline-model-results} for all the three hydrogen lines, and their posterior distributions are shown in Fig. \ref{fig:corner-plot-exponential-fit}. 
It is clear that none of the three previously mentioned conditions is met, suggesting that the exponential model, although able to successfully reproduce the individual profiles separately, fails in simultaneously reproducing them, at least for this specific LRD. 
Our $W_\text{\ha}$ value is consistent with that obtained from single \ha\ modelling, and in general with the results from \citetalias{rusakov2025}, as the exponential widths in their sample vary in the range 1000-2000 \kms.
In contrast, both $W_\text{\hb}$ and $W_\text{\pab}$ are fundamentally unconstrained. 
The Gaussian broad components have widths that differ by a factor of almost 3, and also the scattered fractions are not consistent, nor they satisfy Eq. \ref{eq-relation-fscatt}, as \pab\ exhibits the lowest $f_\text{scatt}$. 

Since our baseline model cannot put satisfactory constraints on $W_{\hb}$ and $W_{\pab}$, we further test the electron scattering scenario by performing additional fits satisfying one of conditions i--iii at a time.  
First, we enforce the same FWHM$_\text{broad}$ for the three hydrogen lines, while allowing the exponential parameters to vary freely (model 2).
Then, we tie both $W$ and $f_\text{scatt}$, and leave FWHM$_\text{broad}$ independent (model 3).  
The best-fit parameters for these models are reported in Table \ref{table-baseline-model-results}.
We also test a more restrictive model where the broad profiles of the three lines are fully tied, i.e. we force them to have same FWHM$_\text{broad}$, $W$ and $f_\text{scatt}$ (model 4).  
Lastly, we compare our set of exponentials with a double-Gaussian model, where broad emission in each line is modelled using two Gaussian components with the same velocity, and independent width. 
For each of our analysed models, both exponential and Gaussian, we report the $\chi^2_\text{red}$ and BIC of the best-fits in Table \ref{table-statistics}. 
Models 2 and 3 perform well, but neither is able to simultaneously satisfy conditions i--iii. 
In model 2, the \hb and \pab exponential profiles exhibit lower widths by a factor 2--4 with respect to the baseline model, and their scattered fractions saturate at $\sim 1$ (Fig. \ref{fig:posteriors-exponential-fit-model2,3}), allowing only for a 3$\sigma$ lower limit estimate. 
Model 3 appears to be the statistically favoured one, with broad line widths differing by a factor of 2 and \ha exhibiting the narrower profile, also evident from Fig. \ref{fig:line-profiles-vel-space}. 
Model 4 yields the poorest fit, particularly for \hb, which shows a broader profile and more pronounced wings than the other two hydrogen lines (Fig. \ref{fig:line-profiles-vel-space}). 
The double Gaussian model performs worse than our exponential models 1--3, likely because the final fit statistic is always dominated by the bright \ha line, which we demonstrate being best reproduced by an exponential, rather than Gaussian, profile (Section \ref{section-emission-line-modelling}), in agreement with previous findings \citepalias{rusakov2025}.  
This strong preference for the exponential model is less pronounced in \hb and \pab, as also indicated by the fact that, for these two lines, the double Gaussian model yields a total broad profile width consistent with that recovered by our baseline exponential model. 

Summarizing, our results rule out the simple scenario in which the central broad lines are scattered by an embedding ionized medium into much more extended exponential wings. This model would indeed predict the same exponential widths and scattered fractions (in absence of dust) for the three hydrogen lines, which are not observed. 
Notably, \cite{juodzbalis2025} also questioned the simple electron scattering scenario on physical grounds. 
Specifically, they argued that an ionized medium surrounding the BLR should recombine and therefore emit in narrower \ha, which should be an order of magnitude more luminous than observed.

\cite{juodzbalis2025} also ruled out the simple Balmer scattering scenario first proposed by \cite{Naidu+2025}, which would also imply black hole mass underestimates by orders of magnitude. 
In this alternative scenario, the \ha and \hb lines are broadened by resonant Balmer scattering, which is expected to redistribute the core flux into the wings, producing double-peaked line profiles, which are not observed in the \textit{Rosetta Stone}, nor in most high-redshift LRDs. 
Furthermore, this model predicts a significantly narrower \pab profile, as this line is not affected (or affected to much lower extend) by Balmer scattring. 
Contrary to these expectations and in agreement with \cite{Juodzbalis2024a}, in our work we find that the \pab has similar width as \hb, and is in fact broader than \ha. 

If not from an ionized medium embedding the BLR, then what is the origin of the broad line wings observed in many high-redshift AGN and LRDs? 
The use of exponential, or Lorentzian, or power-law functional forms (which all differ from the Gaussian profile, but are essentially undistinguishable in the vast majority of spectra) for fitting the broad line wings is not a new proposal -- these profiles have been already largely invoked and used in a variety of previous works for fitting local and lower redshift AGN and quasars \citep{Veron2001_expon,Nagao2006_powerlaw,Mullaney2008_multic_exp,Kollatschny2011_Lorentzrotbroad,Cracco2016_expon,scholtz+2021}. 
However, in the past, the wings of these lower redshift (and much brighter) AGN were mainly explained in terms of macro-turbulence in the BLR, or strong rotation support, or radiation pressure driven winds \citep{Goad2012_mocroturb,Kollatschny2011_Lorentzrotbroad,Baldwin1975}. 
Given our results, these may be simpler explanations which do not imply black hole masses that are two orders of magnitude lower than inferred from the virial relations. Within this context, it is important to note that reverberation mapping of local AGN which display such broad wings that can be fitted with Lorentzian or exponential profiles, follow the same virial relations established for other AGN \citep{Du2019_revmap}.
Finally, recently the black hole mass of a prototypical, lensed LRD at z=7 has been measured directly by resolving its sphere of influence, and the BH mass is found to be fully consistent with the value derived from the virial relations, and about two orders of magnitude higher than inferred by the scattering scenario (Juod{\v z}balis et al. in prep.).

\begin{table}
      \caption[]{Model statistics.}
      \label{table-statistics}
         \centering
         \begin{adjustbox}{width=0.95\linewidth}
         \begin{tabular}{cccc}
         \toprule
         Model & Parameters & $\chi^2_{\text{red}}$ & BIC\\
         \midrule
         Baseline (model 1) & FWHM$_\text{broad}$, $W$, $f_\text{scatt}$ free & 3.08 & 816 \\
         \midrule
         Model 2 & $W$, $f_\text{scatt}$ free & 3.60 & 921\\
         & FWHM$_\text{broad}$ tied &  & \\
         \midrule
         Model 3 & FWHM$_\text{broad}$ free & 3.03 & 797\\
         & $W$, $f_\text{scatt}$ tied &  & \\
         \midrule
         Model 4 & FWHM$_\text{broad}$, $W$, $f_\text{scatt}$ tied & 4.55 & 1118\\
         \midrule
         Double Gaussian & --- & 3.79 & 966\\
         \bottomrule
         \end{tabular}
         \end{adjustbox}
   \end{table}

\section{Summary and Conclusions}
\label{section-conclusions}
In this work, we have tested the hypothesis that the broad \ha emission observed in LRDs is due to electron scattering by an ionized medium embedding the BLR \citepalias{rusakov2025}. 
We have studied the spectral profiles of the \ha, \hb and \pab emission lines in the \emph{Rosetta Stone}, the brightest known LRD (GN-28074 at $z=2.26$), by fitting an exponential model with various setups to all three lines simultaneously. 
While the fit to \ha alone is excellent, these models fail to reproduce all three line profiles simultaneously while satisfying physically motivated constraints on line shapes and scattered fractions (conditions i--iii, Section~\ref{section-discussion}).
Our findings suggest that the electron scattering scenario is either incomplete, potentially relaxing our conditions i--iii, or outright incorrect, even in LRDs like \emph{Rosetta Stone}, where \ha is adequately reproduced. 
We have also ruled out the simple Balmer scattering scenario, based on the finding that Pa$\beta$ is broader than H$\beta$ and has a width similar to H$\alpha$,
as already pointed out by \cite{Juodzbalis2024a}. 
Together, these results establish the \textit{Rosetta Stone} as a clear counterexample to the claim  that black hole masses are universally and systematically overestimated due to electron or Balmer scattering effects.

When comparing models, attention should be paid to the disproportionate weight of \ha in the signal-to-noise ratio distribution. 
More in general, we underscore the importance of
using constraints from multiple lines to robustly investigate the physical
processes at play in AGN.
The claim of black hole masses underestimated by two orders of magnitude in LRDs, because of the putative scattering scenario, is therefore not supported by our results on this prototypical, bright LRD.

\section*{Acknowledgements}
We thank V. Rusakov for his valuable suggestions which improved the clarity and completeness of this work. 
The research activities described in this paper were carried out with contribution of the Next Generation EU funds within the National Recovery and Resilience Plan (PNRR), Mission 4 - Education and Research, Component 2 - From Research to Business (M4C2), Investment Line 3.1 - Strengthening and creation of Research Infrastructures, Project IR0000034 – ``STILES - Strengthening the Italian Leadership in ELT and SKA''. 
MB, FDE, RM, XJ, JS, IJ and GCJ acknowledge support by the Science and Technology Facilities Council (STFC), by the ERC through Advanced Grant 695671 ``QUENCH'', and by the UKRI Frontier Research grant RISEandFALL.
RM also acknowledges funding from a research professorship from the Royal Society.
IJ also acknowledges support by the Huo Family Foundation
through a P.C. Ho PhD Studentship.

\section*{Data Availability}

The data used in this work are public on the JWST archive. 

\bibliographystyle{aa}
\bibliography{bib}

\appendix

\section{Oxygen line fitting and outflow characterisation}
\label{appendix-O3-fit}

\begin{figure}
    \centering
    \includegraphics[width=0.9\linewidth]{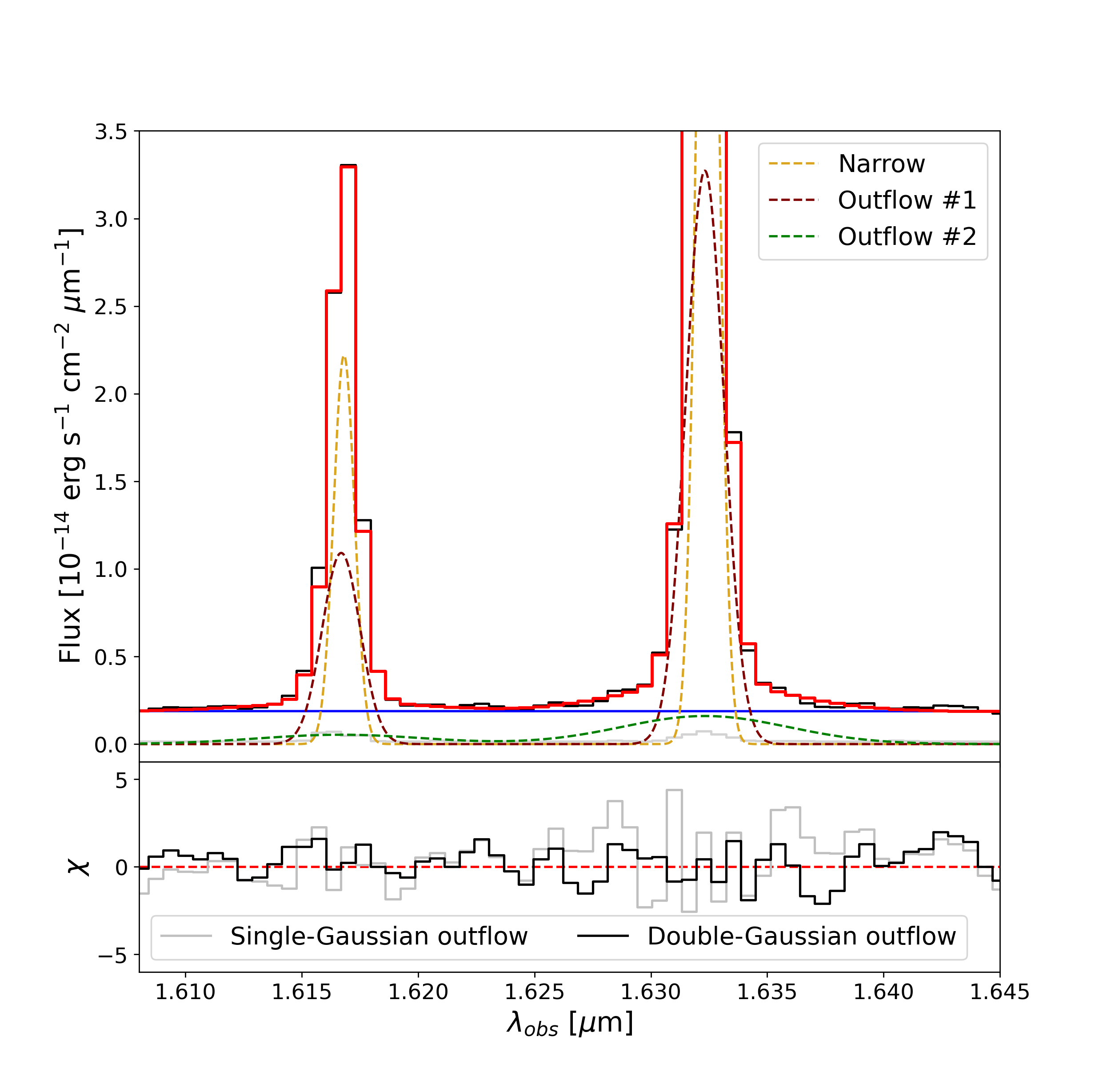}
    \caption{Fit of \oiii$\lambda\lambda 4959,5007$ doublet. Each line is modelled with a narrow and a double-Gaussian outflow components. In the lower panel, we compare the residuals of the fit with the single-Gaussian outflow model, to highlight the inadequacy of the latter in correctly reproducing oxygen emission. }
    \label{fig:O3_fitting_results}
\end{figure}

\begin{table}
\centering
\caption{Line fluxes for the \oiii\ narrow and outflow components, resulting from our double-Gaussian outflow model. They are not dust-corrected.}
\adjustbox{max width=1.0\textwidth}{
\begin{tabular}{cc}
\toprule
 Line & Flux\\[5pt]
 & [$10^{-18}$ erg s$^{-1}$ cm$^{-2}$]\\[5pt]
\midrule
\oiii$_\text{narrow}$ & $80_{-5}^{+3}$\\[6pt]
\oiii$_\text{out,1}$ & $61_{-3}^{+5}$\\[6pt]
\oiii$_\text{out,2}$ & $13_{-1}^{+1}$\\
\bottomrule
\end{tabular}}
\label{table-O3-fluxes}
\end{table}

Before analysing the hydrogen lines, we first fit the \oiii$\lambda\lambda 4959,5007$ doublet separately to identify and, if present, constrain the outflow properties. 
The fitting procedure follows the approach described in Section \ref{section-emission-line-modelling}. 
Each \oiii\ line is modelled with a narrow and a two-Gaussian outflow component. The two outflow line widths are allowed to vary, respectively, in the ranges 100--900 \kms and 900--10,000 \kms, and are constrained to have the same velocity.
Our fitting results are shown in the upper panel of Fig. \ref{fig:O3_fitting_results}.

The need for a double-Gaussian outflow model is motivated by the fact that a single Gaussian outflow model would result in a poor fit, particularly around the brighter \oiii5007 line. 
This is illustrated in the lower panel of Fig. \ref{fig:O3_fitting_results}, where where the residuals from the single- and double-Gaussian fits are shown in gray and black, respectively.
The double-Gaussian outflow model is statistically preferred, with a $\Delta$BIC of 101. 
The two outflow components exhibit a velocity of $v_{\text{out}} = -142 \pm 3$ \kms, and have widths of FWHM$_{\text{out,1}}=302_{-16}^{+12}$ \kms and FWHM$_{\text{out,2}}=1420_{-144}^{+137}$ \kms.
The corresponding line fluxes are summarised in Table \ref{table-O3-fluxes}.
Given that the broader outflow component contributes less than 25\% of the total outflow flux, in the hydrogen line fitting we adopt the width of the narrower outflow component only.

We also attempt an exponential fit of the \oiii\ lines, using the same exponential model described in Section \ref{section-emission-line-modelling}. 
This fit yields to a low values of the scattered fraction ($f_{\text{scatt,\oiii}} \sim 0.1$) and exponential width ($W_\text{\oiii} \sim 490$ \kms), suggesting that oxygen may experience electron scattering only in the outer, lower-density regions of the gas envelope, where oxygen can recombine. 
However, this exponential model does not include an outflow component for \oiii\ --and therefore also for the hydrogen lines--, which is instead necessary for accurately reproducing the latter. 
For this reason, we keep the double-Gaussian model as our fiducial outflow description. 
A more in-depth analysis of outflow properties in GN-28074 is beyond the scope of this work, but will be addressed in future research.

\section{Complete fitting results for the baseline exponential model}
\label{appendix-all-baseline-fitted-params}

In Table \ref{table-tot-fitted-fluxes}, we provide the flux list for all measured hydrogen emission line components: narrow, broad unscattered, broad scattered and outflow. 
The intrinsic narrow FWHM is $89 \pm 3$ \kms, and is the same for the three hydrogen lines, as well as for the \nii$6548,6584$ doublet, which is tentatively detected at 4$\sigma$ confidence. 
We estimate dust extinction for each component of our model from the \ha/\hb\ flux ratio, by assuming a standard case B decrement of 2.86 and the SMC extinction curve from \cite{gordon2003}. 
We obtain $A_V = 1.22_{-0.18}^{+0.24}$ for the narrow, $A_V = 1.05_{-0.39}^{+0.30}$ for the broad unscattered, and $A_V = 5.73_{-0.31}^{+1.00}$ for the broad scattered components, respectively. 
Given that for the \hb\ outflow component our fit provides only an upper limit, we estimate $A_{V, \text{outflow}} > 3$. 

In Table \ref{table-absorption-params} we report the properties of \ha\ and \hb\ absorbers. 
The retrieved covering fraction of $0.98$ suggests that the dense medium is almost entirely covering the BLR. 
Our optical depth estimates differ from those retrieved in \cite{Juodzbalis2024a} because of the different BLR emission model. 
The differences in $\Delta v_{\text{abs}}$ and $\sigma v_{\text{abs}}$ can be attributed to the presence of degeneracies between emission and absorption parameters. 

\begin{table}
\centering
\caption{Line fluxes resulting from our baseline model fitting procedure. They are not dust-corrected.}
\adjustbox{max width=1.0\textwidth}{
\begin{tabular}{cc}
\toprule
 Line & Flux\\[5pt]
 & [$10^{-18}$ erg s$^{-1}$ cm$^{-2}$]\\[5pt]
\midrule
\hb$_\text{narrow}$ & $24_{-1}^{+1}$\\[6pt]
\hb$_\text{broad,unscatt}$ & $30_{-4}^{+3}$\\[6pt]
\hb$_\text{broad,scatt}$ & $10_{-4}^{+3}$\\[6pt]
\hb$_\text{outflow}$ & $<3$\\
\midrule
\ha$_\text{narrow}$ & $108_{-7}^{+7}$\\[6pt]
\ha$_\text{broad,unscatt}$ & $132_{-7}^{+7}$\\[6pt]
\ha$_\text{broad,scatt}$ & $370_{-7}^{+7}$\\[6pt]
\ha$_\text{outflow}$ & $30_{-7}^{+8}$\\[6pt]
\nii6584 & $4_{-0}^{+1}$\\
\midrule
\pab$_\text{narrow}$ & $10_{-1}^{+1}$\\[6pt]
\pab$_\text{broad,unscatt}$ & $70_{-7}^{+3}$\\[6pt]
\pab$_\text{broad,scatt}$ & $<7$\\[6pt]
\pab$_\text{outflow}$ & $13_{-1}^{+1}$\\
\bottomrule
\end{tabular}}
\label{table-tot-fitted-fluxes}
\end{table}

\begin{table}
\centering
\caption{Properties of Balmer absorbers, assuming the same covering fraction, velocity shift and dispersion.}
\adjustbox{max width=1.0\textwidth}{
\begin{tabular}{cc}
\toprule
 Absorption parameter & Value\\
\midrule
$C_f$ & $0.98_{-0.03}^{+0.02}$\\[6pt]
$\tau_{0,\mathrm{H\alpha}}$ & $1.82_{-0.11}^{+0.15}$\\[6pt]
$\tau_{0,\mathrm{H\beta}}$ & $0.73_{-0.08}^{+0.10}$\\[6pt]
$\Delta v_{\mathrm{abs}}$ [\kms] & $-437_{-17}^{+15}$\\[6pt]
FWHM$_{\mathrm{abs}}$ [\kms] & $577_{-20}^{+17}$\\[6pt]
\bottomrule
\end{tabular}}
\label{table-absorption-params}
\end{table}

\setcounter{section}{3}
\renewcommand\thefigure{\Alph{section}\arabic{figure}}
\setcounter{figure}{0}
\begin{figure*}
    \centering
    \includegraphics[width=1\linewidth]{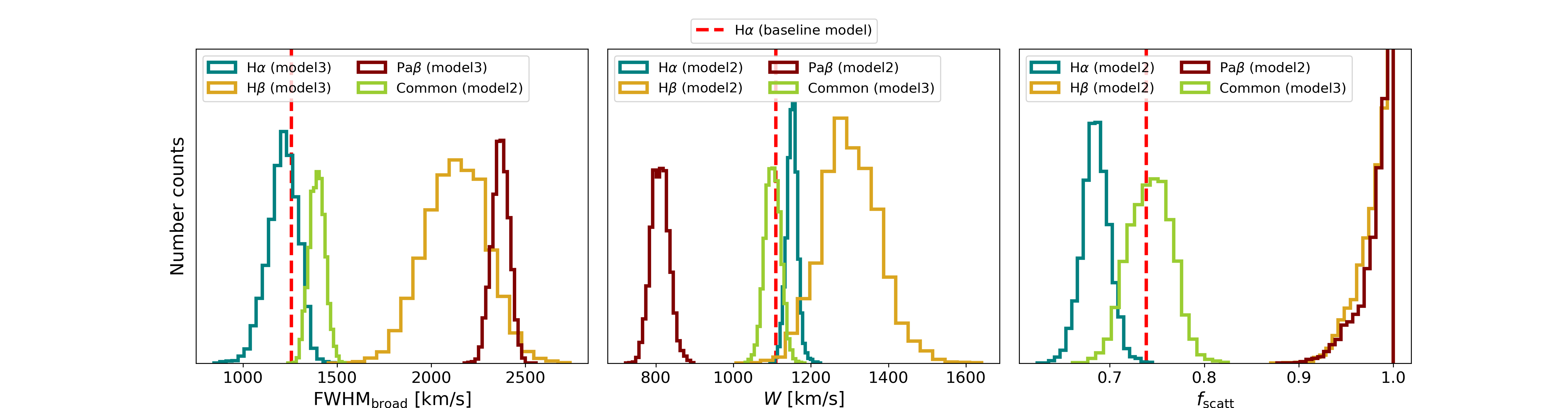}
    \caption{Posterior distributions of FWHM$_{\text{broad}}$, $W$, and $f_\text{scatt}$ from models 2 and 3. In model 2 the three line widths are bound and the exponential parameters are free to vary, while in model 3 it is the opposite. 
    Common parameters are displayed in green, while single-line parameters are displayed in teal for \ha, gold for \hb, and maroon for \pab. }
    \label{fig:posteriors-exponential-fit-model2,3}
\end{figure*}
\setcounter{section}{2}
\section{Posterior distributions of exponential parameters for models 2 and 3}

Fig. \ref{fig:posteriors-exponential-fit-model2,3} displays the posterior distributions of FWHM$_{\text{broad}}$, $W$, and $f_\text{scatt}$ from model 2 (broad widths bound, exponential parameters free) and 3 (broad widths free, exponential parameters bound).

\bsp	
\label{lastpage}
\end{document}